\documentclass[10pt,twocolumn]{IEEEtran}
\usepackage{amssymb}
\usepackage{amsfonts}
\usepackage{amsmath}
\usepackage{algorithm}
\usepackage{algorithmic}
\usepackage{graphicx,subfigure,amsmath,amssymb,cite}

\usepackage[dvips]{color}
\usepackage{float}

\newcommand{\Hnull}{\mathcal{H}_0}
\newcommand{\Halt}{\mathcal{H}_1}

\newcommand{\Honull}{\mathcal{{D}}_0}
\newcommand{\Hoalt}{\mathcal{{D}}_1}

\newtheorem{theorem}{\textbf{Theorem}}

\ifCLASSINFOpdf
\else
\fi
\begin{document}

%=========================================

\title{Delay-Constrained Covert Communications with A Full-Duplex Receiver}
%====================================================Pilot Optimization,
\author{{Feng Shu,~\IEEEmembership{Member,~IEEE,} Tingzhen Xu,~\IEEEmembership{Student Member,~IEEE,}  Jinsong Hu,~\IEEEmembership{Member,~IEEE,} \\and Shihao Yan,~\IEEEmembership{Member,~IEEE,}}

\thanks{F. Shu, T. Xu, and J. Hu are with the School of Electronic and Optical Engineering, Nanjing University of Science and Technology, Nanjing, China. (Emails: \{shufeng, tingzhen.xu, jinsong\_hu\}@njust.edu.cn). F. Shu and J. Hu are also with the College of Physics and Information, Fuzhou
University, Fuzhou 350116, China }

\thanks{S. Yan is with the School of Engineering, Macquarie University, Sydney, NSW, Australia (Email: shihao.yan@mq.edu.au).}

\thanks{This work was supported in part by the National Natural Science Foundation of China under Grant 61771244 and Grant 61472190.
}

}

\vspace{-2cm}

\maketitle
\begin{abstract}
In this work, we consider delay-constrained covert communications  with the aid of a full-duplex (FD) receiver. Without delay constraints, it has been shown that the transmit power of artificial noise (AN) at the FD receiver should be random in order to enhance covert communications. In this work, we show that transmitting AN with a fixed power indeed improve covert communications with delay constraints, since in a limited time period the warden cannot exactly learn its received power. This explicitly shows one benefit of considering practical delay constraints in the context of covert communications. We analyze the optimal transmit power of AN for either fixed or globally optimized transmit power of covert information, based on which we also determine the specific condition under which transmitting AN by the FD receiver can aid covert communications and a larger transmit power of AN always leads to better covert communication performance.
\end{abstract}

\begin{IEEEkeywords}
Covert communication, full-duplex, artificial noise, delay constraint,  transmit power.
\end{IEEEkeywords}
%========================Section I===================================
\section{Introduction}
Covert communication aims to enable a communication between two users while guaranteeing a negligible detection probability of this communication at a warden. It shields the very existence of the transmission and thus mitigates the threat of discovering the presence of the transmitter or communication in wireless networks\cite{Hu2018covertrelay,Hu2018ICC,Liu2018xidian}. As such, it achieves a higher-level security relative to the
conventional information-theoretic secrecy technologies, which only protect the content of transmitted messages. In addition, covert communication can address privacy issues in wireless networks. For example, it can aid to hide a transmitter's location information in Internet of Things (e.g., vehicular networks), where the exposure of location information is a critical privacy concern\cite{Ying2013Vehicular}. As such, covert communication is emerging as a cutting-edge research topic in the context of wireless communication security.

Limits of covert communications over additive white gaussian noise (AWGN) channels was established in \cite{bash2013limits}, which is widely known as the square root scaling law. Covert communications in the context of relay networks was examined in \cite{Hu2018covertrelay}, showing that a relay can opportunistically transmit its own messages to the destination covertly on top of forwarding a source's message. Multi-hop covert communication over an arbitrary network in an AWGN environment and in the presence of multiple collaborating wardens was investigated in~\cite{Azadeh2018Multi}.
Covert communication over the non-fading and the fading channel with a Poisson field of interferers was studied in \cite{Biao2018Poisson}.
%In \cite{Sobers2017Covert}, an external jammer joined the network to generate artificial noise (AN) in order to create uncertainty at the warden Willie to help achieve covert communications.
%We note that the method of transmitting AN by a cooperative external jammer may suffer from some issues, such as mobility and trustworthiness. These issues can be addressed by replacing the external jammer by a full-duplex receiver that can simultaneously receive information signals and transmit AN, which was examined in \cite{Hu2018ICC}.

The aforementioned works studied covert communications under the assumption of asymptotically infinite number of channel uses (i.e., $n\rightarrow\infty$). However, in many application scenarios (e.g., connected vehicles, smart meters, or automated factories  etc.), it requires the transmission of short data packets (e.g., about 100 channel uses), which need to be delivered with stringent requirements in terms of latency~\cite{Makki2015finite}.
 {Against this background, the effect of finite blocklength (i.e., with short delay constraints) on covert communications was examined in \cite{Shihao2018Delay}, which showed that using random transmit power of covert information can further enhance the delay-constrained covert communications.
Although \cite{Shihao2018Delay} examined the impact of finite blocklengh on covert communications, they did not consider sending
AN by a full-duplex (FD) receiver in the context of covert communications. Meanwhile,
\cite{Hu2018ICC} discussed the effect of AN transmitted by a FD receiver  in covert communications with infinite blocklength. As shown in \cite{Hu2018ICC}, the transmit power of AN should be random in order to enable the transmitted AN to benefit the covert communications with infinite blocklength. This is due to the fact that as the number of channel uses approaches infinite, the warden Willie will know the AN power and can cancel its impact by adjusting the detection threshold accordingly \cite{Hu2018ICC}.
In this work, we mainly tackle whether transmitting AN with a fixed power can improve the performance of delay-constrained covert
communications (with finite blocklength) and what are the conditions for achieving the benefit of AN in the context of delay-constrained covert communications.}
%These two facts motivate us to examine the delay-constrained covert communications with a full-duplex receiver, which reveals another benefit brought by the delay constraint to covert communications, i.e., the transmit power of AN can be a fixed value to still enable the transmitted AN to aid the delay-constrained covert communications.
%As studied in \cite{Hu2018ICC}, the full-duplex receiver can use the AN to enhance the performance of the covert transmission. Motivated by these studies, this paper presents an practical approach to the design of optimal power at the FD receiver to improve the performance in the context of delay constraint (i.e., finite blocklength) and draws some insights through our analysis.

%The remainder of this paper is organized as follows. In Section II, we present the system model and adopted assumptions. Then the optimization for the covert transmission is proposed in Section III. The performance of the proposed method is evaluated in Section IV and conclusions are given in Section V.

%\vspace{-0.1cm}
%======================Section II System Model======================
\section{System Model}
\subsection{Adopted Assumptions}
We consider a scenario, where a transmitter (Alice) tries to send messages to a FD receiver (Bob) covertly under the supervision of a warden (Willie), who is detecting whether Alice is communicating with Bob or not. Alice and Willie are assumed to be equipped with a single antenna each, while besides the single receiving antenna, Bob uses an additional antenna to transmit AN to potentially create uncertainty at Willie.

In this work, we assume that the transmit power at Alice denoted by $P_a$ is fixed for all the available channel uses. As per TR 38.802 in 3GPP, the transmission over one channel use only takes roughly 0.01ms \cite{3GPP}. As such, in this work we do not consider different transmit power levels for different channel uses, since varying transmit power within such a short time is not practical.

%In Fig.~\ref{fig:1}, the solid lines represent the covert signal transmission from Alice, while the dotted lines represent the AN signal transmission from Bob.

\subsection{Detection at Willie}
Considering AWGN channels, this work focuses on delay-constrained covert communications, where the number of channel uses $N$ for covert communications is finite and we assume that Alice transmits signals over all the available channel uses\cite{Shihao2018Delay}. The received signal at Willie in the $i$-th channel use is given by
\begin{equation} \label{yw}
{\mathbf{y}_w}[i] =\left\{
\begin{aligned}
&\sqrt {P_b} {\mathbf{v}_b}[i] + {\mathbf{n}_w}[i], &\Hnull,\\
&\sqrt {P_a} \mathbf{x}_a[i] + \sqrt {P_b} {\mathbf{v}_b}[i] + {\mathbf{n}_w}[i], &\Halt,
\end{aligned}
\right.
\end{equation}
where $P_a$ and $P_b$ are transmit power at Alice and Bob, respectively, $\mathbf{x}_a$ is the signal transmitted by Alice satisfying $\mathbb{E}[\mathbf{x}_a[i]\mathbf{x}^{\dag}_a[i]]=1$, $i = 1, 2, \dots, N$, $\mathbf{v}_b$ is the AN transmitted by Bob satisfying $\mathbb{E}[\mathbf{v}_b[i]\mathbf{v}^{\dag}_b[i]]=1$, and $\mathbf{n}_w[i]$ is the AWGN at Willie subject to $\mathbf{n}_w[i] \thicksim\mathcal{CN}(0,\sigma^2_w)$.
The null hypothesis $\Hnull$ means that Alice does not transmit, while $\Halt$ means that Alice transmits messages to Bob.

Based on the received signal in \eqref{yw}, Willie makes a binary decision on whether the received signal comes from $\Hnull$ or $\Halt$. $\Hoalt$ and $\Honull$ represent the binary decisions that Alice transmits or not, respectively.  {In general, the false alarm $\mathbb{P}_{FA}\triangleq \mathbb{P}(\Hoalt|\Hnull)$ and missed detection $\mathbb{P}_{MD}\triangleq \mathbb{P}(\Honull|\Halt)$ are adopted as }the metrics to measure the detection performance at Willie. Accordingly, the optimal test that minimizes the detection error probability $\xi=\mathbb{P}_{FA}+\mathbb{P}_{MD}$ is the likelihood ratio test with $\lambda= 1$ as the threshold  {as following
\begin{align}\label{P0P1}
\frac{\mathbb{P}_1\triangleq\prod_{i=1}^{N}f(\mathbf{y}_w[i]|\Halt)}
{\mathbb{P}_0\triangleq\prod_{i=1}^{N}f(\mathbf{y}_w[i]|\Hnull)}
\mathop{\gtrless}\limits_{\Honull}^{\Hoalt}1.
\end{align}
}
We have a lower bound on $\xi$ according to the Pinsker's inequality\cite{bash2013limits}, which provides us with a theoretical basis for the following analysis and is given by
\begin{align}\label{pinsker}
\xi\geq1-\sqrt{\frac{1}{2}\mathcal{D}(\mathbb{P}_0||\mathbb{P}_1)},
\end{align}
where $\mathcal{D}(\mathbb{P}_0||\mathbb{P}_1)$ is the Kullback-Leibler (KL) divergence from $\mathbb{P}_0$ to $\mathbb{P}_1$, $\mathbb{P}_0$ and $\mathbb{P}_1$ are the likelihood functions under $\Hnull$ and $\Halt$ as per \eqref{P0P1}, respectively.
In this work, we adopt this lower bound as the detection performance metric, since the expressions of $\mathbb{P}_{FA}$ and $\mathbb{P}_{MD}$ are too complicated to be used for further analysis, which is mentioned in\cite{Shihao2018Delay}.
Specifically, $\mathcal{D}(\mathbb{P}_0||\mathbb{P}_1)$ is given by~\cite{Shihao2018Delay}
\begin{align} \label{KL_div}
\mathcal{D}(\mathbb{P}_0||\mathbb{P}_1)=N\left[\ln(1+\gamma_w)-\frac{\gamma_w}{1+\gamma_w}\right],
\end{align}
 {where $\gamma_w=P_a/(\sigma_w^2+P_b)$ is the signal-to-interference-plus-noise ratio (SINR) at Willie. A small value of $\mathcal{D}(\mathbb{P}_0||\mathbb{P}_1)$ means that the distance between $\mathbb{P}_1$ and $\mathbb{P}_0$ is small, which normally leads to a high detection error probability $\xi$ at Willie. }In covert communications, we normally have $\xi \geq 1 - \epsilon$ as the covertness requirement, where $\epsilon$ is an arbitrarily small value. Following \eqref{pinsker}, in this work we adopt $\mathcal{D}(\mathbb{P}_0||\mathbb{P}_1)\leq 2\epsilon^2$ as the covertness requirement.

\vspace{-0.1cm}
\section{Delay-Constrained Covert communications}

In this section, we first determine the optimal transmit power of AN (i.e., $P_b$) at Bob for a fixed feasible $P_a$. Then, we jointly optimize $P_a$ and $P_b$ in order to maximize the performance of covert communications.
\vspace{-0.1cm}
\subsection{Optimization of $P_b$ for a Fixed and Feasible $P_a$}

 For a given transmission rate $R$, the effective throughput $\eta$ can be represented as $\eta=NR(1-\delta)$\cite{Shihao2018Delay}, where $\delta$ is the decoding error probability. For a fixed $R$, this decoding error probability is given by \cite{Shihao2018Delay},
\begin{align}\label{aa}
\delta=Q\left(\frac{\sqrt{N}(1+\gamma_b)\left(\ln{(1+\gamma_b)}+
\frac{1}{2}\ln N-R\ln2\right)}{\sqrt{\gamma_b(\gamma_b+2)}}\right),
\end{align}
where $\gamma_b$ is the SINR at Bob given by
\begin{equation} \label{gamma_b}
\gamma_b=\frac{P_a}{\sigma_b^2+hP_b},
\end{equation}
where $\sigma_b^2$ is the variance of AWGN noise at Bob and $0\leq h\leq1$ is the self-interference cancellation coefficient at Bob corresponding to different cancellation levels \cite{Elsayed2015All,shihao2018fullduplex}.
We denote the entire item in the Q function bracket as $\zeta$. Then, the first derivative of $\zeta$ with respect to $\gamma_b$ is given by
\begin{align}\label{deri_zeta}
\zeta'|_{\gamma_b}\!=\!\frac{\sqrt{N}
\left[\gamma_b^2\!+\!2\gamma_b\!+\!R\ln{2}-\ln{(1\!+\!\gamma_b)}
-\frac{1}{2}\ln N\right]}
{\left[\gamma_b(2\!+\!\gamma_b)\right]^{3/2}}.
\end{align}
The ultimate goal of our design in covert communications is to maximize the effective throughput $\eta$ subject to the covertness constraint and the corresponding optimization problem can be written as
\begin{subequations}\label{P1}
\begin{align}
(\mathbf{P1}) \quad \underset{P_b}{\max} \quad &\eta \\
\text{s. t.} \quad  &\mathcal{D}(\mathbb{P}_0||\mathbb{P}_1)\leq 2\epsilon^2, \label{covertness_cons}\\
&P_b\leq P_b^{\max},\label{power_cons}
\end{align}
\end{subequations}
where $P_b^{\max}$ is the maximum transmit power of AN at the full-duplex Bob. The feasible condition of and the corresponding solution to $\mathbf{P1}$ are presented in the following theorem. Note that the following theorem is valid on the condition that $\gamma_w$ is a small value.

\begin{theorem}\label{theorem1}
 {For given $P_b^{\max}$ and $\epsilon$, the feasible condition of $\mathbf{P1}$ is $P_a \leq P_a^u$, where
\begin{equation} \label{Pa_condition}
P_a^{u}= \frac{\left(P_b^{\max}+\sigma_w^2\right)\left(\epsilon^2+
\sqrt{\epsilon^4+2\epsilon^2N}\right)}{N}.
\end{equation}
With $P_a \leq P_a^u$, the solution to $\mathbf{P1}$ (i.e., the optimal $P_b$) can be approximately achieved as
\begin{equation} \label{Pb_ast}
P_b^{\ast}=\frac{P_aN}{\epsilon^2+\sqrt{\epsilon^4+2\epsilon^2N}}-\sigma_w^2,
\end{equation}
and the maximum effective throughput $\eta^\ast$ to $\mathbf{P1}$ for a given $R$ can be approximately achieved as
\begin{equation}\label{eff_maximum}
\eta^{\ast}=NR(1-\delta^{\ast}),
\end{equation}
where $\delta^{\ast}$ is obtained by substituting $P_b^{\ast}$ into \eqref{aa}.}
\end{theorem}

\begin{IEEEproof}
 {We first note that, as per \eqref{KL_div}, the KL divergence $\mathcal{D}(\mathbb{P}_0||\mathbb{P}_1)$ in the constraint \eqref{covertness_cons} monotonically increases with $P_a$, while it monotonically decreases with $P_b$. As such, considering the  constraint \eqref{power_cons}, the feasible condition of the the optimization problem $\mathbf{P1}$ is in terms of the maximum value of $P_a$ and this maximum $P_a$ is achieved by solving $\mathcal{D}(\mathbb{P}_0||\mathbb{P}_1)= 2\epsilon^2$ with $P_b = P_b^{\max}$.
}

 {In order to derive the explicit expressions for the maximum $P_a$ and the optimal $P_b$, we next detail how to solve $\mathcal{D}(\mathbb{P}_0||\mathbb{P}_1)= 2\epsilon^2$ as a function of $\gamma_w$. Following \eqref{KL_div}, $\mathcal{D}(\mathbb{P}_0||\mathbb{P}_1)= 2\epsilon^2$ can be written as
\begin{align}\label{covert_constraint_eq}
N\left[\ln{(1+\gamma_w)}-\frac{\gamma_w}{1+\gamma_w}\right]=2\epsilon^2.
\end{align}
We note that, in covert communications, $\gamma_w$ is normally very small in order to ensure a high detection error probability at Willie. When $\gamma_w$ is very small, we can adopt the approximation $\ln{(1+x)}\sim x$ in \eqref{covert_constraint_eq}. Applying this, we have
\begin{align}\label{quadratic_eq}
\gamma_w^2-\frac{2\epsilon^2}{N}\gamma_w-\frac{2\epsilon^2}{N}=0.
\end{align}
We note that \eqref{quadratic_eq} is a quadratic equation and there exists two real solutions to it, since the discriminant is given by $4\epsilon^4(1+2N)/N^2>0$. In addition, the product of these two real solutions is $-\frac{2\epsilon^2}{N}<0$, which means that there is always a positive solution to \eqref{quadratic_eq}. Specifically, the two real solutions are given by
\begin{align} \label{gamma_w}
\gamma_{w}^{\pm}=\frac{\epsilon^2\pm\sqrt{\epsilon^4+2\epsilon^2N}}{N}.
\end{align}
We note that $\gamma_{w}^+>0$ and $\gamma_{w}^-<0$ due to $\sqrt{\epsilon^4+2\epsilon^2N}>\epsilon^2$ in \eqref{gamma_w}, which confirms that we always have a positive solution to \eqref{quadratic_eq}, which is $\gamma_{w}^+ = \frac{\epsilon^2+\sqrt{\epsilon^4+2\epsilon^2N}}{N}$.
}

 {
Substituting $\gamma_{w}^+$ into $\gamma_w={P_a}/(\sigma_w^2+P_b)$ and setting $P_b = P_b^{\max}$, we obtain the maximum $P_a$ as given in \eqref{Pa_condition} and $P_a \leq P_a^u$ is the feasible condition of $\mathbf{P1}$. Under this feasible condition, we next derive the optimal $P_b$ for a given feasible $P_a$. To this end, we first note that $\eta$ monotonically decreases with $P_b$, since as proved in \cite{sun2018short-packet}, $\eta$ is a monotonically increasing function of $\gamma_b$ for any valid transmission rate $R$. We also note that $\mathcal{D}(\mathbb{P}_0||\mathbb{P}_1)$ in the constraint \eqref{covertness_cons} monotonically decreases with $P_b$. As such, the optimal $P_b$ is the one that satisfies $\mathcal{D}(\mathbb{P}_0||\mathbb{P}_1)= 2\epsilon^2$. Applying the approximation used to obtain \eqref{quadratic_eq} and noting $\gamma_w={P_a}/(\sigma_w^2+P_b)$, we obtain the optimal $P_b$ as given in \eqref{Pb_ast}. Finally, substituting $P_b^\ast$ into the definition of $\eta$, we achieve the maximum effective throughput  as given in \eqref{eff_maximum}. This completes the proof of Theorem~1.
}
\end{IEEEproof}

\vspace{-0.1cm}
\subsection{Joint Optimization of $P_a$ and $P_b$ }

In this subsection, we jointly optimize $P_a$ and $P_b$ in order to maximize the throughput at Bob in the considered scenario. Specifically,
the focused optimization problem is given by
\begin{equation}\label{P2}
\begin{aligned}
(\mathbf{P2}) \quad \underset{P_a, P_b}{\max} \quad &\eta \\
\text{s. t.} \quad  &\mathcal{D}(\mathbb{P}_0||\mathbb{P}_1)\leq 2\epsilon^2,\\
&P_b\leq P_b^{\max},
\end{aligned}
\end{equation}
where we do not consider a maximum transmit power constraint at Alice, since in covert communications Alice's transmit power is normally low. The solution to this optimization problem $\mathbf{P2}$ is given in the following theorem.

\begin{theorem}\label{theorem2}
 {For any given covertness constraint $\epsilon$ and transmission rate $R$, the maximum $\eta$ to $\mathbf{P2}$ is $\eta^{\ast}=NR(1-\delta^{\ast})$. The corresponding optimal transmit power $P_a$ and AN power $P_b$ in $\delta^{\ast}$ can be approximately achieved as
\begin{align} \label{Pa_ast}
P_a^{\ast}=\frac{\epsilon^2+\sqrt{\epsilon^4+2\epsilon^2N}}{N}
(\sigma_w^2+{P_b^{\ast}}),
\end{align}
\begin{equation} \label{Pb_ast_globel}
P_b^{\ast}=\left\{
\begin{aligned}
&P_b^{\max},&\sigma_b^2\geq h\sigma_w^2,\\
&0,&\sigma_b^2< h\sigma_w^2.
\end{aligned}
\right.
\end{equation}
}
\end{theorem}

\begin{IEEEproof}
 {As per (4) and \eqref{gamma_b}, both $\mathcal{D}(\mathbb{P}_0||\mathbb{P}_1)$ and $\eta$ monotonically increase with $P_a$. As such, the equality in the constraint $\mathcal{D}(\mathbb{P}_0||\mathbb{P}_1)\leq 2\epsilon^2$ is always guaranteed by an optimal $P_a$ for any fixed $P_b$. We can prove this by contradiction. We first suppose that the optimal $P_a$, denoted by $P_a^\ddag$, is achieved with $\mathcal{D}(\mathbb{P}_0||\mathbb{P}_1)< 2\epsilon^2$. Since both the KL divergence $\mathcal{D}(P_0||P_1)$ in the constraint $\mathcal{D}(\mathbb{P}_0||\mathbb{P}_1) \leq  2\epsilon^2$ and the objective function $\eta$ monotonically increase with $P_a$, we can still increase $P_a^\ddag$ in order to improve $\eta$ while still ensuring $\mathcal{D}(\mathbb{P}_0||\mathbb{P}_1)\leq 2\epsilon^2$.
 This contradicts the supposition that $P_a^\ddag$ is the optimal $P_a$. Therefore, $\mathcal{D}(\mathbb{P}_0||\mathbb{P}_1)= 2\epsilon^2$ is guaranteed by the actually optimal $P_a$ (denoted by $P_a^\dag$) for any given $P_b$.} This leads to the fact that the optimal $P_a$ can be expressed as a function of $P_b$, which can be achieved following a similar approach as used in the proof of Theorem~1. This expression is the same as \eqref{Pa_ast}, where we have to replace $P_a^\ast$ and $P_b^\ast$ with $P_a^\dag$ and $P_b$, respectively. Since $\eta$ is a monotonically increasing function of $\gamma_b$ for given transmission rate $R$ and $N$ \cite{sun2018short-packet}, we next tackle the monotonicity of $N\gamma_b$ to clarify the monotonicity of $\eta$. Then, substituting $P_a^{\dag}$ into $N\gamma_b$, we have
\begin{align}\label{Ngammab}
 (N\gamma_b)^{\dag}={\frac{(\epsilon^2+\sqrt{\epsilon^4+2\epsilon^2N})
 (\sigma_w^2+P_b)}
{\sigma_b^2+hP_b}}.
\end{align}
We derive the first derivative of $ (N\gamma_b)^{\dag}$ with respect to $P_b$ as
\begin{align}\label{first_derivative}
\frac{\partial (N\gamma_b)^{\dag}}{\partial P_b}=\frac{(\epsilon^2+\sqrt{\epsilon^4+2\epsilon^2N})
 (\sigma_b^2-h\sigma_w^2)}{(\sigma_b^2+hP_b)^2}.
\end{align}
As per \eqref{first_derivative}, we can see that the sign of it is solely determined by the term $(\sigma_b^2-h\sigma_w^2)$.
 {Then we have the following two cases:}

 {\emph{\textbf{Case 1}:} When $\sigma_b^2 < h\sigma_w^2$, $(N\gamma_b)^{\dag}$ {(and the corresponding $\eta$)} decreases with $P_b$ and then we have $P_b^{\ast}=0$ to maximize the effective throughput $\eta$.}

 {\emph{\textbf{Case 2}:} When $\sigma_b^2 \geq h\sigma_w^2$, $(N\gamma_b)^{\dag}$ {(and the corresponding $\eta$)} increases with $P_b$ and then we have $P_b^{\ast}=P_b^{\max}$ to maximize the effective throughput $\eta$.}

 {Considering these two cases above, we achieve the desired results in \eqref{Pa_ast} and \eqref{Pb_ast_globel}, which completes the proof.}
\end{IEEEproof}
Note that $\mathcal{D}(\mathbb{P}_1||\mathbb{P}_0)$ is also a widely used metric in covert communications. Similar to the analysis of $\mathcal{D}(\mathbb{P}_0||\mathbb{P}_1)$, we can still draw the conclusions that when $\sigma_b^2\geq h\sigma_w^2$, $P_b^{\ast}=P_b^{\max}$, and when $\sigma_b^2< h\sigma_w^2$, $P_b^{\ast}=0$.

 {This result first indicates that transmitting AN with fixed power can still benefit the delay-constrained covert communications, when we have $P_b^{\ast}=P_b^{\max}$. Considering the self-interference at the full-duplex receiver, it is also reasonable to observe $P_b^{\ast}=0$ in a specific case, which demonstrates that transmitting AN does not help covert communications under this case.}

\vspace{-0.1cm}
\section{Simulations and Discussions}

\begin{figure}[t]
\centering
\includegraphics[width=3.5in, height=2.8in]{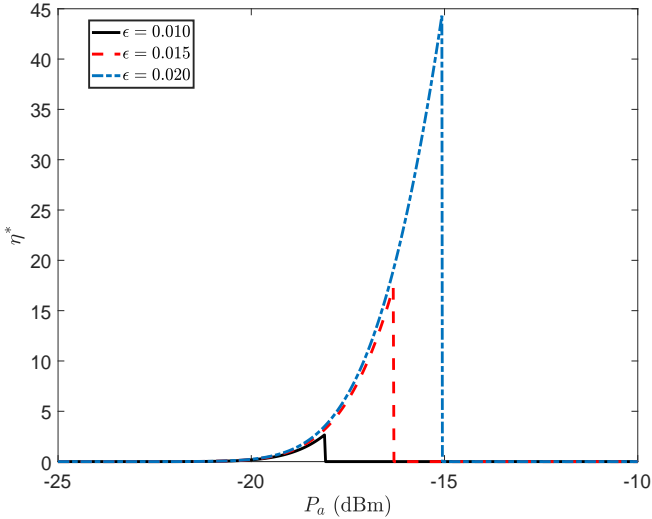}\\
\caption{$\eta^{\ast}$ versus $P_a$, where $P_b^{\max}=1 \textmd{dBm}$, $\sigma_b^2=\sigma_w^2=0~\textmd{dBm}$, $R=3.4$, $N=100$, and $h=0.01$.}\label{fig:1}
\vspace{-0.2cm}
\end{figure}

\begin{figure}[t]
\centering
\includegraphics[width=3.5in, height=2.8in]{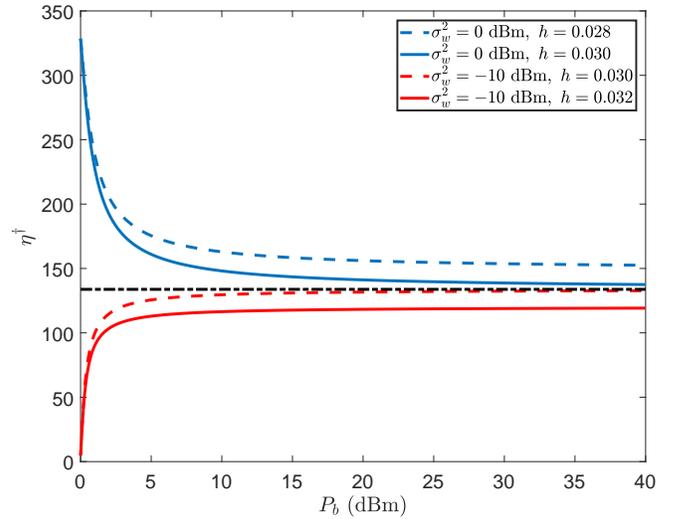}\\
\caption{$\eta^{\dag}$ versus $P_b$, where $\sigma_b^2=-20$~dBm, $\epsilon=0.01$, $R=3.4$, $N=100$, and $P_b^{\max}=40$~dBm.}\label{fig:2}
\vspace{-0.2cm}
\end{figure}

In Fig.~\ref{fig:1}, we plot the maximum $\eta$, i.e.,  $\eta^{\ast}$, achieved by the optimal $P_b$ for a fixed $P_a$ with different values of $\epsilon$.
We first observe that $\eta^{\ast}$ increases with $P_a$ before all the turning points, which can be explained by substituting $P_b^\ast$ given in \eqref{Pb_ast} into \eqref{gamma_b} and noting $\sigma_b^2 > h \sigma_w^2$ in the settings of Fig.~1. We note that
the covertness constraint cannot be satisfied when $P_a$ exceeds a certain value due to the maximum transmit power constraint at Bob, which can explain why we have
$\eta^{\ast}=0$ when $P_a$ is larger than some specific values. Furthermore, we observe that an increase in $\epsilon$ relaxes the covertness constraint and leads to a higher value of $\eta^{\ast}$.
In Fig.~\ref{fig:2}, we plot the maximum $\eta$, i.e., $\eta^{\dag}$, achieved by the optimal $P_a$ and $P_b$ with different values of $\sigma_w^2$ and $h$.
The two blue curves are obtained in Case~1, i.e., when $\sigma_b^2<h\sigma_w^2$, where as confirmed we have $P_b^\ast=0$.
Meanwhile, the two red curves are achieved for Case~2, i.e., when $\sigma_b^2\geq h\sigma_w^2$, where we have $P_b^{\ast}=P_b^{\max}$.
In this figure, we also observe that, for the red curves, the $\eta^\dag$ decreases with $h$, which shows that a more efficient self-interference cancellation can improve covert communications. Finally, we observe that, for the red curves,  $\eta^\dag$ tends to an upper bound as $P_b$ increases.

\vspace{-0.1cm}
%================================
\section{Conclusions}
In this work, we studied covert communications with delay constraints over AWGN channels with the aid of a FD receiver. Specifically, we examined the possibility and strategy of using the FD receiver to transmit AN in order to shield the covert transmission from Alice to Bob. Our examination shows that a fixed AN transmit power can improve delay-constrained overt communications. In addition, the conducted analysis indicates that in most practical scenarios the transmit power of AN should be as large as possible when Alice's transmit power can be jointly optimized. We also determine the specific condition under which transmitting AN by the FD receiver can aid covert communications and a larger transmit power of AN always leads to better covert communication performance.

\bibliographystyle{IEEEtran}
\bibliography{IEEEabrv,CC}

\end{document}